\documentclass[letter,fleqn,usenatbib]{mnras}

\usepackage{newtxtext,newtxmath}

\usepackage[T1]{fontenc}
\usepackage{ae,aecompl}

\usepackage{graphicx}	
\usepackage{amsmath}	
\usepackage{amssymb}	

\title[Hunting for Brown Dwarfs in M\,4]{Hunting for Brown Dwarfs in the Globular Cluster M4: 2nd epoch HST NIR observations}

\author[A. Dieball et al.]{
A. Dieball$^{1}$\thanks{E-mail: adieball@astro.uni-bonn.de},
L. R. Bedin$^{2}$,
C. Knigge$^{3}$,
M. Geffert$^{1}$,
R. M. Rich$^{4}$,
A. Dotter$^{5}$,
\newauthor
H. Richer$^{6}$,
D. Zurek$^{7}$
\\
$^{1}$Argelander Institute for Astronomy, University of Bonn, Auf dem H\"ugel 71, 53121 Bonn, Germany\\
$^{2}$INAF - Osservatorio Astronomico di Padova, Vicolo dell'Osservatorio 5, I-35122 Padova, Italy\\
$^{3}$Physics and Astronomy, University of Southampton, SO17 1BJ, UK\\
$^{4}$Department of Physics and Astronomy, University of California at Los Angeles, Los Angeles, CA 90095-1547, USA\\
$^{5}$Harvard-Smithsonian Center for Astrophysics, 60 Garden Street, Cambridge, MA 02138, USA\\
$^{6}$Department of Physics and Astronomy, University of British Columbia, Vancouver, BC, V6T 1Z1, Canada\\
$^{7}$Department of Astrophysics, American Museum of Natural History, New York, NY 10024, USA
}

\date{Accepted XXX. Received YYY; in original form ZZZ}

\pubyear{2017}

\begin{document}
\label{firstpage}
\pagerange{\pageref{firstpage}--\pageref{lastpage}}
\maketitle

\begin{abstract}
  We present an analysis of the second epoch HST WFC3 $F110W$
  near-Infrared (NIR) imaging data of the globular cluster M\,4. The
  new dataset suggests that one of the previously suggested four brown
  dwarf candidates in this cluster is indeed a high-probability
  cluster member. The position of this object in the NIR colour
  magnitude diagrams (CMDs) is in the white dwarf/brown dwarf
  area. The source is too faint to be a low-mass main sequence star,
  but, according to theoretical considerations, also most likely
  somewhat too bright to be a bona-fide brown dwarf. Since we know
  that the source is a cluster member, we determined a new optical
  magnitude estimate at the position the source should have in the
  optical image. This new estimate places the source closer to the
  white dwarf sequence in the optical-NIR CMD and suggests that it
  might be a very cool ($T_{eff} \le 4500$ K) white dwarf at the
  bottom of the white dwarf cooling sequence in M\,4, or a white
  dwarf/brown dwarf binary. We cannot entirely exclude the possibility
  that the source is a very massive, bright brown dwarf, or a very
  low-mass main sequence star, however, we conclude that we still have
  not convincingly detected a brown dwarf in a globular cluster, but
  we expect to be very close to the start of the brown dwarf cooling
  sequence in this cluster. We also note that the main sequence ends
  at $F110W\approx 22.5$ mag in the proper-motion cleaned CMDs, where
  completeness is still high.
\end{abstract}


\section{Introduction}

Brown dwarfs are astronomical objects that are not massive enough to
sustain hydrogen fusion. As such, they are ``failed'' stars, and
indeed present a link between planets and stars
\citep{kulkarni1997}. Brown dwarfs were first postulated half a
century ago \citep{kumar1963,hayashi1963}, but the first detection of
a brown dwarf was reported only 30 years later in 1995
\citep{nakajima1995, rebolo1995}. Only in the last decade, large
numbers of brown dwarfs have been detected, thanks to increased
sensitivities and field of views of modern instruments which enabled
large surveys like the Two Micron All Sky Survey
\citep{skrutskie2006}, the Sloan Digital Sky Survey \citep[SDSS,
][]{york2000}, the United Kingdom Infrared Telescope Deep Sky Survey
\citep{lawrence2007}, and the Wide-field Infrared Survey Explorer
\citep{wright2010}. However, due to the intrinsic faintness of these
sources, the detection of brown dwarfs is strongly biased towards
younger (and hence brighter) and metal-rich brown dwarfs in the solar
neighbourhood and in young star clusters or star forming regions
\citep[e.g. ][]{casewell2014, boudreault2013}. In contrast, we know of
only very few old and metal-poor brown dwarfs \citep[and lowest-mass
main sequence stars, see e.g. ][]{burgasser2003, burgasser2009,
  lepine2004, burgasser2006, cushing2009, burningham2014, troup2016}.

Globular clusters, on the other hand, are among the most metal-poor
and oldest stellar aggregates in our Galaxy. They offer the advantage
that all members of a globular cluster share the same distance and (to
first order) the same age and metallicity - properties that are
difficult to obtain for field sources. Because of dynamical evolution,
we might expect that most low-mass stars and brown dwarfs would
evaporate and be lost to the host globular cluster that have ages of
10 Gyr or more. However, by virtue of their high mass and stellar
density, globular clusters potentially host still large numbers of
brown dwarfs. This is based not only on their large number of stars
but also, stellar interactions might have increased the numbers of
brown dwarfs in these systems \citep[e.g. ][]{thies2010, thies2015,
  stamatellos2011, kaplan2012}.

However, no brown dwarfs have been identified in any globular cluster
to date. Since brown dwarfs cannot sustain hydrogen fusion, they
become cooler and fainter with time. Thus, we expect that the brown
dwarfs in the old Galactic globular clusters are fainter than the
lowest mass stars on the main sequence (MS), which sustain their
luminosity via hydrogen burning for longer than a Hubble time. Because
brown dwarfs in globular clusters have cooled for typically more than
10 Gyr, as a further result of brown dwarf cooling, we indeed expect a
gap in the colour-magnitude diagrams (CMDs) of globular clusters
between brown dwarfs and the end of the MS. Identifying such faint,
sub-stellar objects in globular clusters is therefor challenging and
restricts this kind of research to the globular clusters closest to
us. The globular cluster M\,4 (NGC\,6121) is one of the closest
($\approx2$ kpc, e.g. Bedin et al. 2009) and has been subject to
ultra-deep optical studies with the Hubble Space Telescope (HST), see
\citet{richer1997, richer2004, hansen2004, bedin2009} that reach to
the bottom of the white dwarf cooling sequence, but just fall short of
reaching the end of the MS.

In a previous paper, we reported the detection of four faint sources
in the globular cluster M\,4 \citep{dieball2016}, based on deep
near-infrared (NIR) HST/Wide Field Camera 3 (WFC3) data. Our NIR
photometry reaches beyond the expected end of the hydrogen-burning
limit toward fainter sources and into the white dwarf/brown dwarf
area. Since the white dwarf and the brown dwarf cooling sequence seem
to cross in the NIR CMD, we cannot distinguish between white dwarfs
and brown dwarfs based on the NIR CMD alone. Archival HST optical data
were used for proper-motion cleaning of the CMD and to segregate the
white dwarfs from brown dwarf candidates. Since the optical data are
not deep enough to detect brown dwarfs, any faint NIR sources lacking
optical counterparts are plausible brown dwarf candidates. We found in
total four such sources. However, these four sources are only brown
dwarf {\it candidates}, since prior to this study we had no constraint
on their membership in the cluster.

In this paper, we present the second epoch NIR HST WFC3 imaging data
that are used to proper motion clean all of the first epoch NIR data,
and to find cluster members among our brown dwarf candidates. In
Sect.~\ref{data}, we describe the observations and data reduction,
followed by an analysis of the results in Sect.~\ref{cmd} and our
conclusions in Sect.~\ref{conclusion}.

\section{Data and Data Reduction}
\label{data}

The observations were obtained with the HST WFC3 IR $F110W$
camera/filter combination for one orbit each on 27th March and 25th
July 2017 (program GO-14725, two orbits of $F110W$ imaging, PI:
A. Dieball). A total of eight exposures were obtained, each with an
exposure time of 653 sec, resulting in a total exposure time of 5224
sec. We observed the same field as in the first NIR epoch (GO-12602,
two orbits of $F110W$ imaging and four orbits of $F160W$ imaging), but
used a modified four-point WFC3-IRDITHER-BOX-MIN dither pattern and a
sampling of NSAMP\,14 SPARS\,50 to optimise both the sampled
point-spread function (PSF) as well as the available exposure time.

A geometrically corrected master image was created based on the
pipeline-produced flat-fielded (FLT) images, using {\tt tweakreg} and
{\tt astrodrizzle} running under {\tt PyRAF} (the Python-based
interface to {\tt IRAF}\footnote{{\tt IRAF} (Image Reduction and
  Analysis Facility) is distributed by the National Astronomy and
  Optical Observatory, which is operated by AURA, Inc., under
  cooperative agreement with the National Science Foundation.}). Note
that the dithered imaging data allowed us to refine the pixel scale of
the master image\footnote{We used $final\_scale = 0.0898$ and
  $final\_pixfrac = 0.9$ for the master image.}. This master image
serves as reference image to the photometry software {\tt DOLPHOT}
\citep{dolphin2000}, which runs on the individual flat-fielded images
and provides source positions and magnitudes calibrated to the VEGAmag
system. We started with the parameter setting that we used for the
first NIR epoch data, and then refined some parameters to push our
photometry as deep as possible and maximise the detection of faint
sources as well as keeping spurious detections somewhat at
bay.\footnote{We used Force1 = 1, FlagMask = 4 to eliminate saturated
  stars, WFC3IRpsfType = 1 for the Anderson PSF cores
  \citep{anderson16}, FitSky = 2, SigFind = 1.5, SigFindMult = 0.95,
  SigFinal = 1.5, and RPSF = 10, and Align = 4 as the reference image
  has a smaller resolution than the input flat-fielded images. We then
  selected the {\tt DOLPHOT} photometry for the following matching and
  plotting to an object type of no more than 2 (i.e. only ``stars''),
  a sharpness between -0.5 and +0.5, and a crowding of $\le 0.5$. Note
  that these selection criteria are less strict than the best-fit
  selection for the 1st epoch NIR CMD, see Fig.~4 in
  \cite{dieball2016}. However, we use this approach to push to fainter
  sources and thus an even deeper photometry.} This resulted in 15762
detections, but we caution that these also contain a large number of
spurious detections. For more details on the data reduction, creating
the master image, and the photometry, see also \cite{dieball2016}.

\subsection{Artificial Star Tests}
\label{AStests}

In order to asses the error on the photometry as well as the
completeness in our data sets, we carried out artificial star (AS)
tests using {\tt dolphot}. We use the same master and input images and
parameter setting as in the corresponding original {\tt dolphot}
photometry, and rerun {\tt DOLPHOT} again with the fakestar parameter
set. We expect that not every artificial star will be recovered, which
might be due to the artificial source being placed too close to the
image edge, or on bad (flagged) pixels, or too close to or on a bright
star. Also, the magnitude {\tt DOLPHOT} derives is not exactly
identical to the input magnitude, which is also expected as
environmental effects (blending, crowding, background) impact the
derived magnitudes. All these effects are expected to depend on the
input magnitude, i.e. a faint star is less likely to be recovered and
with a larger deviation from its input magnitude than a bright
star. Thus, artificial star experiments are powerful tests of the
quality of the photometry.

In order to increase the statistics, a large number of artificial
stars should be used. However, this will considerably increase the run
time of the AS photometry. Also, since the photometry is performed for
each star on each individual input image, the run time also depends on
the number of images used. Thus, we ran several sets of AS tests to
increase the statistics, but limited the number of stars per run to
decrease the run time. For the 2nd epoch of $F110W$ observations,
which comprised only eight exposures, we used 100,000 artificial stars
in total. For the 1st epoch of NIR observations, which consists of
eight $F110W$ and sixteen $F160W$ exposures, we used a smaller number
of about 15,000 artificial stars to keep the run time at bay.

\begin{figure*}
  \includegraphics[width=\textwidth]{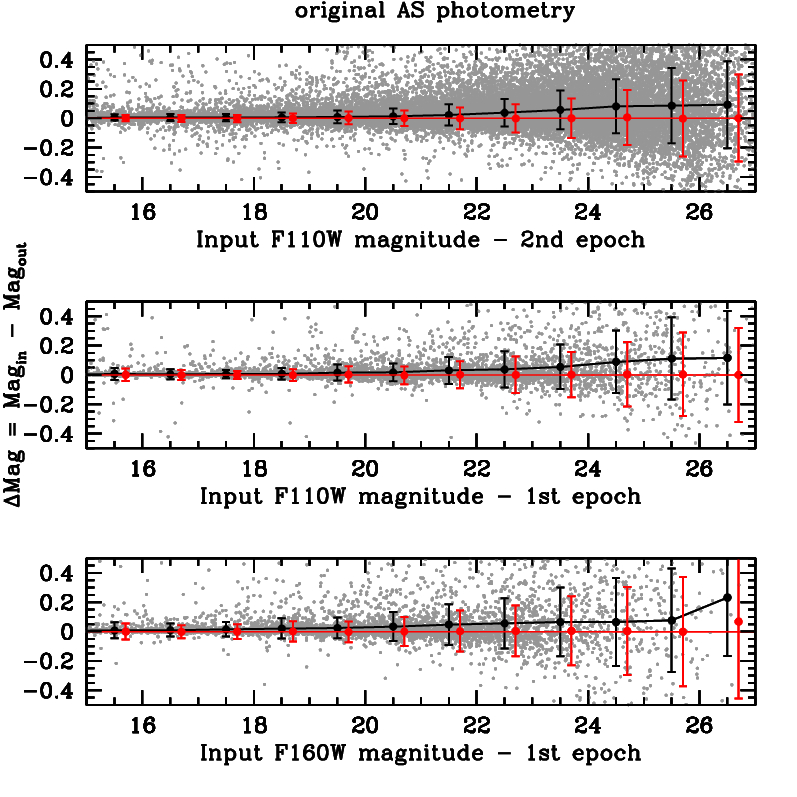}
  \caption{Difference between input and obtained magnitude for
    artificial stars (dark grey data points). The photometries were
    performed using dolphot and the same parameter set as the
    respective original photometry. The data points with error bars
    denote the mean differences $\overline{\Delta Mag} =
    \overline{Mag_{in} - Mag_{out}}$ and their standard deviations per
    magnitude bin for the 2nd epoch $F110W$ data (top panel), the 1st
    epoch $F110W$ (middle panel), and the $F160W$ data (bottom
    panel). A clear offset towards fainter magnitudes can be seen in
    all filters and all epochs. Note that we only include artificial
    sources within a 0.5 pixel tolerance radius from the input
    coordinates and within 0.75 magnitude difference. The red data
    points and error bars denote the mean and corresponding standard
    deviations after photometric correction. To guide the eye, we have
    marked $\Delta Mag =0$ with a red line. See the text for more
    details.}
  \label{meanerror}
\end{figure*}

Note that we applied the same selection criteria on the artificial
source photometry as on the real source photometry, i.e. we allowed
only for object type of no more than 2, a sharpness between $-0.5$ and
$+0.5$, and a crowding of $\le 0.5$. For our error analysis, we use
only AS sources that have a magnitude difference of no more than 0.75
magnitudes and are within a 0.5 pixel tolerance radius from the input
coordinates. Fig.~\ref{meanerror} shows the difference between input
and output magnitude versus input magnitude for the selected AS data
set.

We divided the magnitude range covered by our AS photometry into bins
of one magnitude, and calculated the mean of the differences,
determined as $\overline{\Delta Mag} = \overline{Mag_{in} -
  Mag_{out}}$, per magnitude bin, plotted as black data points with
the corresponding standard deviation as error bars, and listed in
Table~\ref{dselmeanerror}. As can be seen, the mean differences are
very small, but their standard deviations increase towards fainter
magnitudes. Ideally, the mean difference should be zero, but it
becomes larger towards fainter magnitudes, i.e. the recovered
magnitude is on average brighter than the input magnitude. This
suggests that fainter sources are systematically overestimated and are
actually even somewhat fainter than what our photometry derived. In
order to correct for this bias, we used the mean differences as
fiducial points for a polynomial fit, and applied the obtained
correction to all stars in both the AS and the real data
photometry. The red data points in Fig.~\ref{meanerror} denote the
means and standard deviations of the corrected AS photometry, see also
Table~\ref{dselmeanerror}.  The correction worked very well for the
1st and 2nd epoch $F110W$ data, but we caution that a small offset
remained for the faintest magnitude bin that includes only a small
number of sources fainter than 26 mag in the $F160W$ data. However,
this is close to the detection limit anyway, and because the error at
this limit is larger than this small offset, we did not attempt
further correction.

\begin{figure*}
  \includegraphics[width=\textwidth]{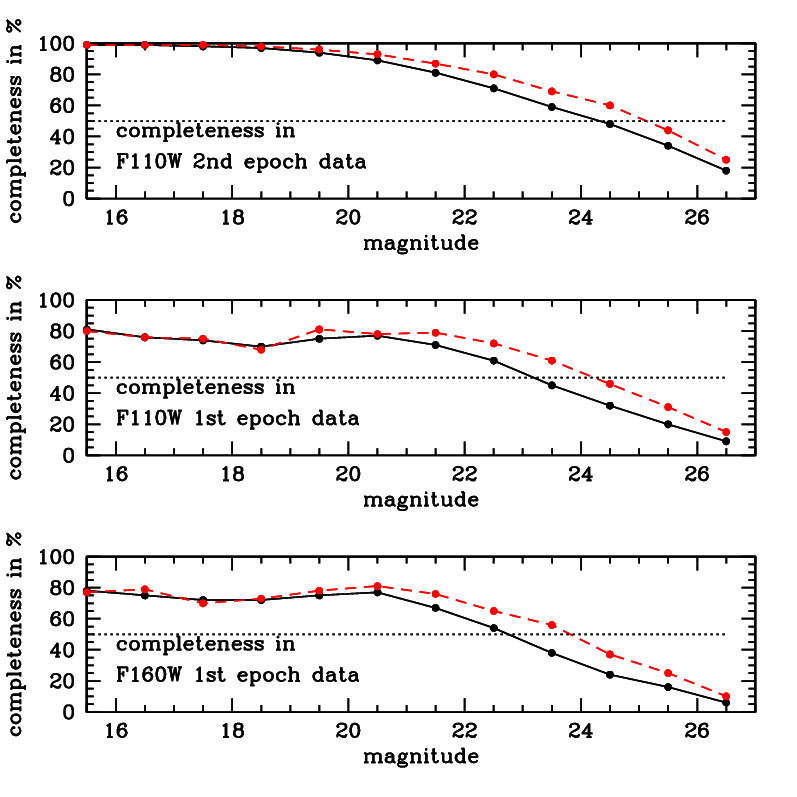}
  \caption{Completeness in \% based on artificial star tests. The
    bottom panel shows the completeness in the (only available) first
    epoch $F160W$ data, the middle of the first epoch $F110W$ data,
    and the top panel of the second epoch $F110W$ data. We distinguish
    between the completeness in the entire master image (black line)
    and completeness outside a $\approx 2.4\arcmin$ radius around the
    cluster centre that contains all four brown dwarf candidates (red dashed
    line). As expected, the completeness for fainter sources (fainter
    than 22 mag) is somewhat higher outside the $\approx 2.4\arcmin$
    radius because this area is less crowded and contains fewer bright
    stars. See the text for more details.}
  \label{completeness}
\end{figure*}

Most artificial sources were recovered by our {\tt DOLPHOT}
routine. Again, we expect the percentage of artificial sources that
are recovered to be magnitude dependent, i.e. fainter sources are less
likely to be found and with a larger magnitude error, see
above. Fig.~\ref{completeness} shows the percentage of AS sources
recovered that fulfil our photometric selection, per magnitude bin for
each filter and epoch. As expected, the completeness drops towards
fainter magnitudes and is at 50~\% around magnitudes of
$F110W_{2nd}\approx 24.5$ mag, i.e. just around the expected end of
the H-burning sequence \cite[see the discussion in Sect.3.3 in
][]{dieball2016}. In the 1st epoch NIR data set, the completeness
drops to 50~\% around $F110W_{1st}\approx 23$ mag and
$F160W\approx 23$ mag, about one magnitude above the end of the
MS. The difference in completeness between the 1st and 2nd observing
epoch is likely due to differences in the sky background and due to
different dither patterns used (we used a standard four-point
WFC3-IRDITHER-BOX-MIN dither pattern for the first epoch, but modified
the dither pattern for the second epoch so that spikes from bright
stars are at different positions in the eight exposures).

Whether a source is detected or not depends on its environment,
i.e. sky background, crowding and nearby bright stars and their PSF
halos and streaks. A faint source is therefore less likely to be
detected in the crowded inner parts of a globular cluster - and with a
larger photometric error. Thus, we compare the completeness of the
entire image (black line in Fig.~\ref{completeness}) with the
completeness of the area in our master images that is outside a radius
of $\approx 2.4\arcmin$ from the cluster centre, i.e. the region on
the master images outside two core radii (red line in
Fig.~\ref{completeness}).  This area includes all former four brown
dwarf candidates. As expected, the completeness in this outer area is
somewhat higher at fainter magnitudes and reaches 50~\% at
$F110W_{1st}\approx 24.5$, $F160W\approx 24$ and $F110W_{2nd}\approx
25$ mag.

\begin{table*}
  \centering
  \caption{Mean differences, $\overline{\Delta Mag} = \overline{Mag_{in} - Mag_{out}}$, 
    and standard deviations for all NIR filters and epochs, based 
    on artificial star experiments. Bin sizes are 1 magnitude, and 
    we list the middle of the bin in the first column, followed by 
    the mean differences per magnitude bin in the 2nd epoch $F110W$ 
    (2nd column), the 1st epoch $F110W$ (3rd column), and the $F160W$
    data. Only sources with less than
    0.75 magnitude difference and that are found within 0.5 pixel 
    tolerance radius of the input coordinates are considered.
    Lines starting with * denote the mean and standard deviation 
    after photometric correction.}
  \label{dselmeanerror}
\begin{tabular}{cccc}
mag  & $\overline{\Delta F110_{2nd}}\pm \sigma_{F110_{2nd}}$ & $\overline{\Delta F110_{1st}}\pm \sigma_{F110_{1st}}$ & $\overline{\Delta F160}\pm \sigma_{F160}$ \\
     & [mag]       & [mag]        & [mag]  \\
\hline
15.5 & $0.00\pm0.02$ & $0.01\pm0.04$ & $0.01\pm0.05$ \\
*    & $0.00\pm0.02$ & $0.00\pm0.04$ & $0.00\pm0.05$ \\
16.5 & $0.00\pm0.02$ & $0.00\pm0.03$ & $0.01\pm0.04$ \\
*    & $0.00\pm0.02$ & $0.00\pm0.03$ & $0.00\pm0.04$ \\
17.5 & $0.00\pm0.03$ & $0.01\pm0.03$ & $0.02\pm0.05$ \\
*    & $0.00\pm0.03$ & $0.00\pm0.03$ & $0.00\pm0.05$ \\
18.5 & $0.01\pm0.03$ & $0.01\pm0.04$ & $0.02\pm0.07$ \\
*    & $0.00\pm0.03$ & $0.00\pm0.04$ & $0.00\pm0.07$ \\
19.5 & $0.01\pm0.04$ & $0.02\pm0.06$ & $0.02\pm0.07$ \\
*    & $0.00\pm0.04$ & $0.00\pm0.06$ & $0.00\pm0.07$ \\
20.5 & $0.01\pm0.05$ & $0.02\pm0.06$ & $0.03\pm0.10$ \\
*    & $0.00\pm0.05$ & $0.00\pm0.06$ & $0.00\pm0.10$ \\
21.5 & $0.02\pm0.07$ & $0.03\pm0.09$ & $0.05\pm0.14$ \\
*    & $0.00\pm0.07$ & $0.00\pm0.09$ & $0.00\pm0.14$ \\
22.5 & $0.04\pm0.09$ & $0.04\pm0.12$ & $0.06\pm0.17$ \\
*    & $0.00\pm0.10$ & $0.00\pm0.12$ & $0.00\pm0.17$ \\
23.5 & $0.06\pm0.13$ & $0.06\pm0.15$ & $0.07\pm0.23$ \\
*    & $0.00\pm0.13$ & $0.00\pm0.16$ & $0.01\pm0.24$ \\
24.5 & $0.08\pm0.18$ & $0.09\pm0.21$ & $0.07\pm0.30$ \\
*    & $0.01\pm0.19$ & $0.00\pm0.22$ & $0.00\pm0.30$ \\
25.5 & $0.09\pm0.26$ & $0.11\pm0.28$ & $0.08\pm0.35$ \\
*    & $0.00\pm0.26$ & $0.00\pm0.28$ & $0.00\pm0.37$ \\
26.5 & $0.09\pm0.30$ & $0.12\pm0.32$ & $0.23\pm0.40$ \\
*    & $0.00\pm0.30$ & $0.00\pm0.32$ & $0.07\pm0.52$ \\
\end{tabular}
\end{table*}

\section{The proper motion cleaned CMDs}
\label{cmd}

The 2nd epoch photometry was matched to the first epoch NIR
photometry, using 166 known cluster sources with magnitudes $15 \le
F110W \le 20$ mag that can be clearly identified in both 1st and 2nd
epoch master images. The 2nd epoch coordinates were transformed to the
1st epoch image pixel coordinates using the tasks {\tt geomap} and
{\tt geoxytran} running under {\tt PyRAF}.  We allowed for up to 2
WFC3 IR pixel matching tolerance between the two epochs. The resulting
vector displacement diagrams (VDDs) and the corresponding CMDs are
plotted in Figs.~\ref{pmcmdIR1ep12} and \ref{pmcmdIR2ep12}. The top
panel in Fig.~\ref{pmcmdIR1ep12} shows the VDDs between the first and
second epoch of $F110W$ data. Because we used known cluster members
for the coordinate transformation, we expect the cluster stars to be
located around $\Delta X = 0$ and $\Delta Y = 0$, and indeed we see a
tight cluster of data points at that location. A second accumulation
of data points can be seen around $\Delta X \approx -0.4$ and $\Delta
Y \approx 0.6$, which is mostly due to Galactic bulge stars, see
\citet{bedin2003}. The second row of VDDs shows the displacement
between the 1st NIR epoch and and the optical $F775W$ dataset. These
VDDs are the same as the VDDs shown in Figs. 5 and 6 in
\citet{dieball2016}. The bottom panel shows the corresponding CMDs for
sources with {\it both an optical and a second epoch $F110W$
  counterpart}, plus the four brown dwarf candidates (which have no
optical counterpart, see Dieball et al. 2016).

We marked our previous four brown dwarf candidates with green (not a
cluster member) and red (cluster member) data points, and as can be
seen, all four appear in both $F110W$ epochs, and hence in the top VDD
panel. All magnitudes and the displacements for all four brown dwarf
candidates are listed in Table~\ref{bds}. Three of the brown dwarf
candidates show a displacement larger than 0.1 WFC3 IR pixels, and
thus are not considered to be cluster members. Indeed, their location
in the VDDs agrees more with field stars. However, our former brown
dwarf candidate BD2 shows a displacement of just 0.05 WFC3 IR pixels,
which places it in the centre of the VDD that is occupied by cluster
members.

\begin{table}
  \centering
  \caption{Brown dwarf candidates detected in the first NIR epoch. The second to 
    fifth columns list the magnitudes measured in the first epoch $F110W$, 
    the second epoch $F110W$, and the $F160W$ and images, followed 
    by the displacement in the pixel coordinates compared to the 
    first epoch $F110W$ master image. The magnitude uncertainties are the 
    errors returned from the {\tt DOLPHOT} routine.}
  \label{bds}
  \begin{tabular}{ccccccc}
    ID & $F110W_{1st}$ & $F110W_{2nd}$ & $F160W$ & shift\\
       & [mag]       & [mag]        & [mag]  & pixel\\
    \hline
    1  & 24.27$\pm$0.02 & 24.24$\pm$0.02 & 23.39$\pm$0.02& 0.58\\
    2  & 25.41$\pm$0.05 & 25.71$\pm$0.06 & 24.75$\pm$0.06& 0.05\\
    3  & 24.36$\pm$0.02 & 24.28$\pm$0.02 & 23.60$\pm$0.02& 0.85\\
    4  & 26.75$\pm$0.16 & 26.80$\pm$0.16 & 26.13$\pm$0.18& 1.01\\
    \hline
  \end{tabular}
\end{table}

\subsection{Cluster membership}

The position of BD2 in the VDDs, with a displacement of only 0.05 WFC3
IR pixel between the first and second NIR epochs, already suggests
that this source is indeed a cluster member.

However, the VDD is also occupied by members of the Galactic bulge and
by field stars. Could BD2 be a member of the field or the bulge
population?  Fig.~\ref{cmdvdd} shows the CMDs and corresponding VDDs
split into seven magnitude bins with widths of 2 mags. Note that we
show only sources that are recovered in all three observing epochs:
the optical, and the first and the second NIR epochs. The black data
points in the CMDs denote sources that show a displacement of no more
than 0.1 pixel in {\it both} sets of VDDs (black data points in the
VDDs). Sources with a larger displacement are shown in grey in the
VDDs, and sources with a displacement of no more than 0.2 pixel in
either set of the VDDs are also plotted as grey data points in the
CMDs. The position of BD2 is marked with a red dot in the CMDs.

Interestingly, the bulge population starts to show up only for sources
fainter than $F110W< 17$ mag, and recedes again for sources fainter
than 25 mag. On the other hand, the cluster sources are prominent
throughout the whole magnitude range. We also notice that the
distribution of cluster sources fainter than 25 mag is broader,
i.e. they show a larger displacement up to 0.3 pixel (but we
conservatively only plot sources with a displacement of up to 0.2
pixel as grey data points in the CMDs, see above). Since bulge sources
predominantly appear at magnitudes brighter than $F110W < 25$ mag, we
might assume that BD2 is not a bulge member.

\begin{figure*}
  \includegraphics[width=\textwidth]{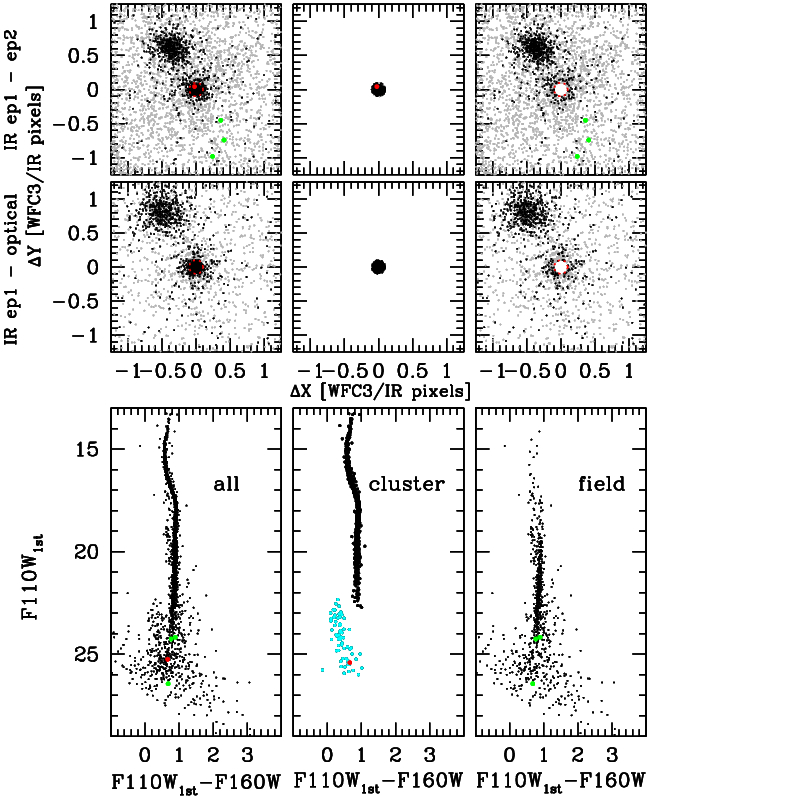}
  \caption{Top row: Vector displacement diagrams (VDDs) between the
    1st and 2nd NIR observing epochs. Middle row: VDDs between the
    optical and the 1st NIR data. Sources that appear in {\it all}
    observing epochs are plotted in black, sources that appear only in
    the corresponding VDD are plotted in grey. Bottom row: The
    resulting NIR CMDs. From left to right: all data; only sources
    that have a displacement of less than 0.1 WFC3 pixel in all VDDs
    which suggests that the sources are cluster members; the remaining
    field sources that have displacements larger than 0.1 WFC3 IR
    pixel. The white dwarfs are selected from the optical-NIR CMDs
    \citep[see][]{dieball2016} and are plotted in cyan. Note that in
    the CMDs we plot 1st epoch NIR data only, but only sources that
    also have a counterpart in the second $F110W$ observing epoch,
    i.e. sources that appear in the optical, the first and the second
    $F110W$ observing epochs (black data points in the VDDs). We also
    add the previously reported brown dwarf candidates, plotted in green, if
    their displacement agrees with being a field source, or in red, if
    the displacement is less than 0.1 WFC3 IR pixel.}
  \label{pmcmdIR1ep12}
\end{figure*}

\begin{figure*}
  \includegraphics[width=\textwidth]{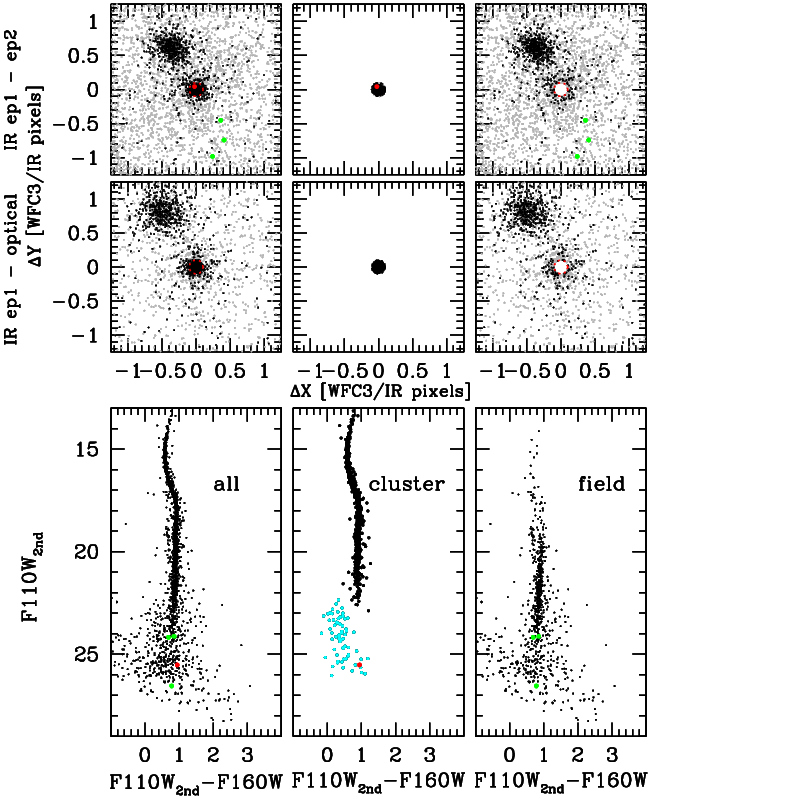}
  \caption{Same as Fig.~\ref{pmcmdIR1ep12}, but the CMDs show the
    second epoch $F110W$ data against the $F160W$ data. (For the
    latter, we only have a first epoch, no second epoch of $F160W$
    data were obtained.)}
  \label{pmcmdIR2ep12}
\end{figure*}

But how many bulge and field sources can we expect at the position of
M\,4 in the VDDs, and hence in the CMDs? We use a simplistic approach:
To get an estimate for the number of field stars we assume that the
distribution of field stars is uniform across the VDD. Thus, we can
simply count the number of field stars and scale this number to the
area occupied by the cluster in the VDD. In total, we count 245
sources outside a 0.2 pixel radius centred at 0,0 (i.e. the cluster)
and outside a 0.4 pixel radius centred at $-0.35$, $0.6$ (the bulge)
in the NIR VDD (i.e. top panels in Figs.~\ref{pmcmdIR1ep12} and
\ref{pmcmdIR2ep12}). We caution that we only consider sources that
also have an optical counterpart (black data points). Out of those,
only 60 sources are within a magnitude range of $25 < F110W < 26$ mag
(the magnitude range in which we find BD2). Scaled to the area of the
cluster, we can expect in total 0.6 field sources, and only 0.15 field
sources within $25 < F110W < 26$ mag.

If we assume the same for the bulge, we find 464 sources within a
circular region centred at $-0.35$, $0.6$ and with a radius of 0.4
pixel in the NIR VDD, 23 of those are in the magnitude range $25 <
F110W < 26$ mag. Scaled to the cluster, we would expect 29 bulge
sources in total, and 1.4 sources with $25 < F110W < 26$ mag. However,
the bulge population is clearly not uniformly distributed in the
VDD. We assume a normal distribution, and fitting a Gauss function to
the bulge distribution in $\Delta X$ and $\Delta Y$ (NIR VDD), we
determine a mean (and hence centre of the bulge distribution in the
NIR VDD) of $\Delta X = -0.36$ and $\Delta Y = 0.61$, and $\sigma =
0.12$. For sources in the magnitude range of BD2, i.e. $25 < F110W <
26$ mag, the centre of the distribution is slightly shifted and
broader at $\Delta X = -0.39$ and $\Delta y = 0.67$ with a $\sigma =
0.18$. A 0.4 pixel radius corresponds to $3.3 \sigma$ and encompasses
99.9\% of the total bulge population. Thus, only 0.46 sources are
expected outside the 0.4 pixel radius. At the position of the cluster
and hence BD2 in the VDDs ($6\sigma$), the number of bulged sources
that can be expected is $4.6\times10^{-4}$.

We conclude that although it is certainly not impossible that BD2 is a
bulge or a field star, it appears very unlikely considering the number
of bulge or field sources that are expected in that area in the VDD,
especially at such faint magnitudes.

To determine membership probability, we used the method described in
\citet{sanders}. The method uses two bivariate Gaussian distributions
(for field and cluster stars) to fit the measured proper motion
distributions in right ascension and declination. From the ratio of
the two distributions the probability of the membership can be
determined for each proper motion. In order to determine the
probability of the source BD2 to be a member of M\,4, we converted the
positional differences shown in Figs.\ref{pmcmdIR1ep12} and
\ref{pmcmdIR2ep12} into proper motions. Only the data from the 1st and
2nd NIR epoch were taken, which provides an epoch difference of 5.1
years (since the source does not appear in the archival optical
observations, we could not use the optical data as a further
epoch). The VDDs in Figs.\ref{pmcmdIR1ep12} and \ref{pmcmdIR2ep12}
corresponds to the classical vector-point-plot diagram. Since the
diagrams in Figs.\ref{pmcmdIR1ep12} and \ref{pmcmdIR2ep12} show a very
clear separation of the cluster and the field stars, we would expect a
high probability of BD2 belonging to M\,4. Our calculations resulted
in a membership of more than 99\%, which is in good agreement with the
results above.

\section{Discussion and Conclusions}
\label{conclusion}

\begin{figure*}
  \includegraphics[width=\textwidth]{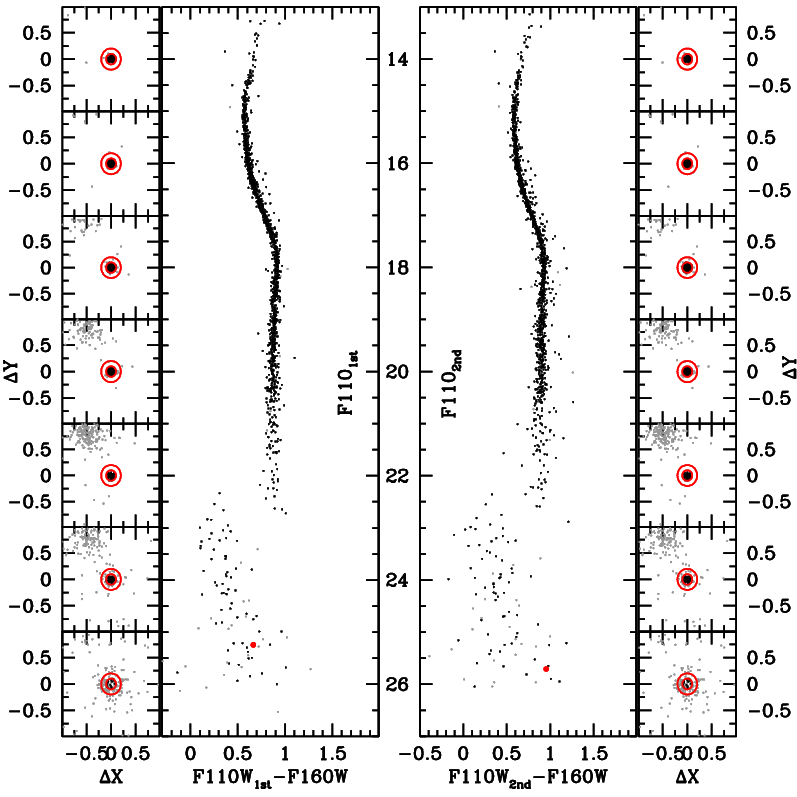}
  \caption{CMDs and corresponding VDDs (outer left and outer right
    panels), for the 1st epoch NIR data (first two panels), and the
    second epoch NIR data (last two panels, from left to right). We
    show the VDDs for sources in magnitude bins of 2 mags in
    $F110_{st}$ and $F110_{2nd}$. The sources marked in black in the
    VDDs are only cluster sources, i.e. sources that show a
    displacement of less than 0.1 pixel between the epochs. Note that
    we only show sources that are recovered in {\it all} three epochs:
    the optical, the first and second NIR epoch. Grey data points in
    the CMDs denote sources that have a displacement between 0.1 and
    0.2 pixel, i.e. sources located in the red annulus in the
    VDDs. BD2, which has no optical counterpart, is marked with a red
    data point. See the text for details.}
  \label{cmdvdd}
\end{figure*}

\begin{figure*}
  \includegraphics[width=\textwidth]{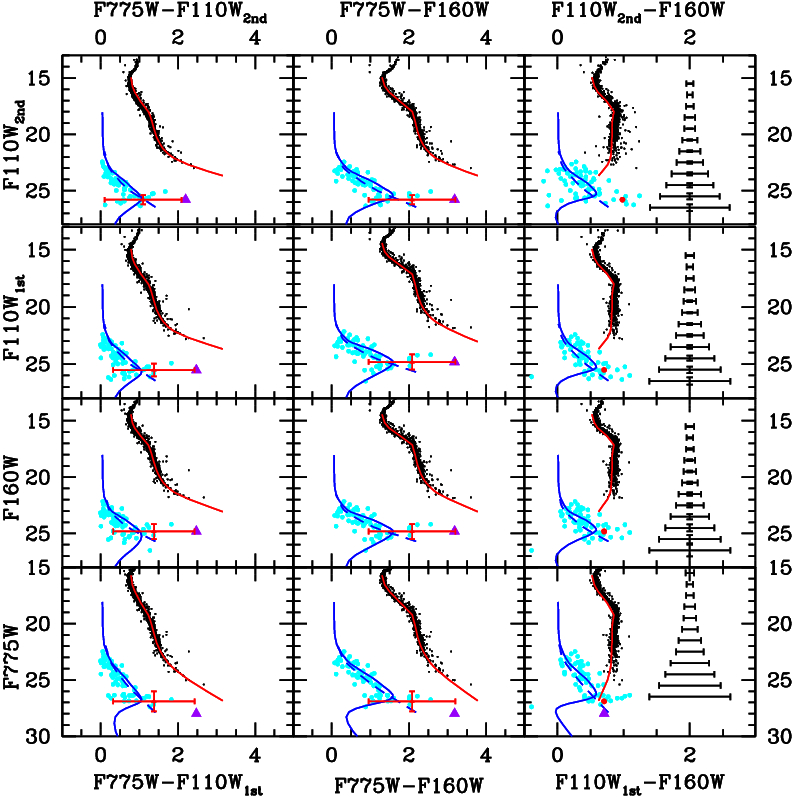}
  \caption{CMDs in all filters and all epochs, we only show sources
    that have a displacement of no more than 0.1 WFC3 pixel between
    the optical, the first and the second NIR observing epochs, which
    suggests that they are cluster members. The blue solid line
    denotes a white dwarf cooling sequences (for masses of 0.5
    M$_\odot$), the blue dashed line a He white dwarf cooling
    sequence, and the red solid line a 12 Gyr BT-Settl isochrone. All
    models fit the underlying data well. Mean photometric errors have
    been obtained using AS tests and are plotted on the right side of
    the NIR CMDs, see also Fig.~\ref{meanerror} and
    Sect.~\ref{AStests}. Note that the mean error was determined for 1
    mag bins, and the error on the colour depends on the magnitude (or
    rather magnitude bin) in question. For the sake of simplicity,
    they are determined only for $F110-F160=0$ mag but have been
    shifted to the right hand side of the CMDs for visibility. Our
    brown dwarf candidate BD2 is marked with a red dot in the NIR
    CMDs. The violet triangle denotes the position of BD2 in the
    optical based on our old estimate \citep{dieball2016}, the red
    cross denotes its position based on the comparison of aperture to
    PSF photometry, see Sect.~\ref{conclusion} for more details.}
  \label{allcmdsep12}
\end{figure*}

Our NIR observations are designed to push into the the brown dwarf
region of the CMDs. But is BD2 a brown dwarf? Since we do not have an
optical measurement, it is very difficult to classify this source.
Fig.~\ref{allcmdsep12} shows the CMDs in all filters for the first and
second NIR epochs. Note that we only show sources that have
measurements in all epochs, and overplot a 12 Gyr BT-Settl isochrone
\citep[red line, ][]{allard1997, allard2013} with a metallicity of
$\rm{M}/\rm{H} = -1$ dex, and a white dwarf (solid blue line) and a He
white dwarf (dashed blue line) cooling sequence\footnote{The white
  dwarf cooling sequences were kindly provided by Pierre Bergeron in
  the HST WFC3 IR filters. We assumed white dwarf masses of 0.5
  $\rm{M}_\odot$. See
  (http://www.astro.umontreal.ca/~bergeron/CoolingModels.}
\citep{holberg, kowalski, tremblay, bergeron}, all scaled to a
distance modulus of 11.2 mag and a reddening $\rm{A}_{\rm{V}} = 1.5$
which gives a reasonably good fit to the underlying data. The
errorbars given on the right side of the NIR CMDs are derived from our
AS photometry, see Sect.~\ref{AStests}. We caution again that we could
not get a {\it measurement} for BD2 in the archival optical
data. Thus, we show its {\it previously estimated} position in the
optical-NIR CMDs \citep[see][]{dieball2016} as a violet triangle,
which is on the red side of the bottom of the white dwarf sequence,
but on the blue side of the MS, i.e. between the sequences. We had
thus classified BD2 as a white dwarf/brown dwarf candidate. Indeed,
based on our previous optical estimate, BD2 appears to be too faint
and too red to be classified with confidence as a white dwarf, but
might just be faint enough to be one of the brightest cluster brown
dwarfs, but possibly too blue. However, the location of the brown
dwarf sequence in {\bf old and metal-poor} stellar populations is so
far unknown. Theoretical considerations \citep[e.g. ]{baraffe1998,
  allard1997} suggest that the brown dwarf sequence turns to the blue
and crosses the white dwarf sequence in the NIR. Indeed, this is what
we see in the NIR CMDs in Figs.~\ref{pmcmdIR1ep12} to
\ref{allcmdsep12}.

But what type of source is BD2?  

{\bf Could BD2 be a very low-mass MS star?} We estimated the end of
the H-burning sequence at $F110W\approx24$ mag. However, the H-burning
limit is highly uncertain, see the discussion in \citet{dieball2016},
and depends on the mass of the lowest-mass star that can still support
nuclear fusion, which, in turn, depends on the metallicity of the
source \citep[see also Fig.~8 in][]{dieball2016}. Previous
theoretical work \citep[e.g.][]{kumar1963, burrows1993} suggested that
the H-burning limit is at higher masses for more metal-poor stars,
probably already at 0.09 M$_\odot$. If true, this also places the
H-burning limit at a brighter magnitude, possibly brighter than our
estimate. This agrees nicely with the best-fit NIR CMD presented in
\citet[][their Fig. 5]{dieball2016}, which shows a well populated MS
down to $F110W\approx24$ mag, which then peters out until it crosses
the white dwarf sequence around $F110W\le25$, when source numbers
increase again.

BD2 is at $F110W\approx25.5$ mag and thus is fainter than our
estimated H-burning limit. It seems unlikely that BD2 is a low-mass MS
star. However, since this limit is uncertain, if BD2 would indeed be a
MS source, it would be the lowest-mass MS star ever detected in a
globular clustery to date. If true, this would seriously challenge our
understanding of the H-burning limit, as it would imply that the
lowest-mass stars that can still sustain hydrogen burning in old (> 10
Gyr) populations would be fainter and at lower masses than previously
thought.

{\bf Could BD2 be a brown dwarf?} We consider it unlikely that BD2 is
a MS source, because it is fainter than the expected end of the
H-burning MS. But is BD2 a brown dwarf? Brown dwarfs cannot sustain H
burning, and as a result, they become cooler and fainter with
time. \cite{caiazzo} presented MESA models for 11 to 13 Gyr old
stellar populations, and predict a gap between the end of the MS and
the beginning of the brown dwarf cooling sequence. For a 12 Gyr
cluster, \citet{caiazzo} predicts this gap to be 3 magnitudes wide
(see their Fig.~3 and Sect.~4). However, these calculations were done
for solar metallicity and in the NIR filters of the James Webb Space
Telescope. As discussed above, the metallicity has an impact on the
faint end of the H-burning sequence, and likely on the shape and the
start of the brown dwarf cooling sequence. In a metal-poor cluster
with an age of $\approx 12$ Gyr, like M\,4, the start of the brown
dwarf cooling sequence is predicted to be at $F110W = 26$ mag and
fainter (Caiazzo, private communication). Thus, Caiazzo's et al.\
(2017) theoretical considerations seem to suggest that BD2 might be
too bright to be a bona fide brown dwarf.

\begin{figure*}
  \includegraphics[width=\textwidth]{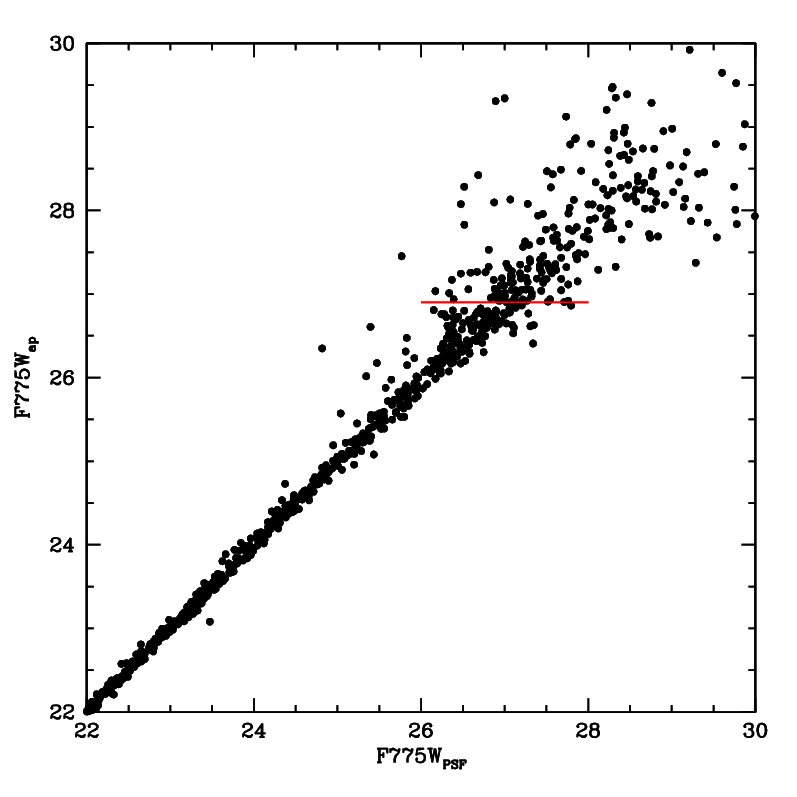}
  \caption{PSF photometry vs. aperture photometry for sources fainter
    than $F775W > 22$ mag. The spread between PSF and aperture
    photometry becomes larger for fainter source. The estimated
    magnitude for BD2 is marked with a red line.}
  \label{apphot}
\end{figure*}

{\bf Could BD2 be a white dwarf?} Since the white dwarf and brown
dwarf cooling sequences cross in the NIR, it is not possible to
distinguish white dwarfs from brown dwarfs based on the NIR CMD alone,
not even after removing field objects via proper motions. An optical
counterpart to BD2 could not be measured, however, a very faint smudge
can be seen in the optical master image at the position of BD2.
The coolest white dwarfs in M4 have temperatures around 4000K
\citep{bedin2009}, just below the blue hook (see
Fig.~\ref{allcmdsep12}). Based on our previous optical estimate, BD2
appears fainter and redder than the blue hook at the bottom of the
white dwarf cooling sequence (usually seen as the endpoint of the
white dwarf cooling sequence) (although it might still be consistent
with the white dwarf sequence considering the large errors at such
faint magnitudes). However, we now know that BD2 is a cluster member,
which allows us to do better on the optical estimate: We worked out
the position BD2 would have in the optical image if it would not have
moved with respect to the other cluster members. This might introduce
a very small error, as also cluster stars do move, but since the
change in position is expected to be less than 0.1 WFC3 pixel, we
expect this error to be small indeed. Next, we run aperture photometry
using {\tt DAOPHOT} \citep{stetson} on all known source positions in
the optical master image, and compared the aperture photometry to the
PSF-fitting photometry to work out the offset between these two types
of photometry, see Fig.~\ref{apphot}. This step is necessary because
PSF-fitting photometry does not work on the position of BD2 in the
optical image as the source signal, if any, is too faint. As can be
seen, the spread between the aperture and PSF-fitting photometry
becomes larger for fainter sources. Based on this, we now estimate the
optical magnitude of BD2 at $F110W \approx 27 \pm 1$ mag. Note the
large uncertainty on this estimate. Indeed we find faint sources whose
aperture-based estimate deviates by more than 2 magnitudes from the
PSF-fitting measurement. We marked the position of BD2 based on the
{\it new optical estimate} with a red cross in
Fig.~\ref{allcmdsep12}. Our new estimate places BD2 closer to the blue
hook on the white dwarf sequence in the optical-NIR CMDs. Based on
this, we conclude that BD2 is likely one of the optically faintest and
coolest white dwarfs detected in this cluster. BD2's NIR colour and
$F110W$ magnitude suggest $T_{eff} \le 4500$ K for a white dwarf with
a pure hydrogen atmosphere, or $T_{eff} \le 2750$ K for a He white
dwarf.

We conclude that BD2 is a high-probability cluster member, and, based
on the information that we have gathered so far, is most likely not a
single brown dwarf but rather a white dwarf. BD2 could well be an
exotic source like a He white dwarf (see Fig.~\ref{allcmdsep12}), or a
binary system consisting of both a (He) white dwarf and a bright and
massive brown dwarf which might explain its red colour compared to
single faint white dwarfs. The source is too faint to be a bona-fide
MS star, and likely too bright (and hence too blue in the optical-NIR
colour) to be a bona-fide brown dwarf. However, we still caution that
we cannot strictly exclude the possibility that BD2 is in fact a very
low-mass MS star - and if so, the faintest, coolest and lowest-mass MS
star ever detected in a globular cluster to date, or a very bright and
massive brown dwarf.

Future NIR observations that include medium wavebands like F127M and
F139M might help to distinguish white dwarfs from brown dwarf
candidates. This is because the spectral energy distributions (SEDs)
of brown dwarfs is governed by molecular absorption of H$_{2}$O in the
IR, which shifts flux to the NIR wavebands (e.g. Allard et al. 1997),
whereas we expect the SED of white dwarfs to be mostly flat.

We also note the sharp drop in the luminosity function of the MS at
$F110W\approx 22.5$ mag. The proper-motion cleaned CMDs in
Figs.~\ref{pmcmdIR1ep12}, \ref{pmcmdIR2ep12}, \ref{cmdvdd} and
\ref{allcmdsep12} suggest that the MS ends around $F110W\approx 22.5$
mag and $F110W - F160W\approx 1$ mag where the completeness is still
high at these magnitudes (around 70\% (entire field) and 80\% (outer
region) in the second $F110W$ epoch, and around 60\% to 70\% (entire
field) and 70\% to 80\% (outer region) in the first NIR epoch, see
Fig.~\ref{completeness}), and the magnitude uncertainty is only around
a tenth of a magnitude. This result might put important constraints on
theoretical models on star formation and evolution.

\section*{Acknowledgements}

AD acknowledges support from the German Federal Ministry of Economics
and Technology (BMWi) provided through DLR under project
50\,OR\,1707. We thank Pierre Bergeron for providing the white dwarf
models (http://www.astro.umontreal.ca/~bergeron/CoolingModels) in the
WFC3 IR filters, and Ingo Thiess and Ilaria Caiazzo for useful
discussions. RMR acknowledges financial support from grants GO-14711
and 14725 from NASA through the Space Telescope Science Institute.

\bsp	
\label{lastpage}

\begin{thebibliography}{99}
\bibitem[\protect\citeauthoryear{Alessi \& Pudritz}{2018}]{alessi2018}
  Alessi, M. \& Pudritz, R. E. 2018, MNRAS accepted
\bibitem[\protect\citeauthoryear{Allard et al.}{1997}]{allard1997}
  Allard, F., Hauschildt, P. H., Alexander, D. R. \& Starrfield,
  S. 1997, ARA\&A, 35, 137
\bibitem[\protect\citeauthoryear{Allard et al.}{2013}]{allard2013}
  Allard, F., Homeier, D., Freytag, B., Schaffenberger, W. \&
  Rajpurohit, A. S. 2013, MSAIS, 24, 128
\bibitem[\protect\citeauthoryear{Anderson}{2016}]{anderson16}
  Anderson, J. 2016, Instrument Science Report WFC3 2016-12, Empirical
  Models for the WFC3/IR PSF
\bibitem[\protect\citeauthoryear{Baraffe et al.}{1998}]{baraffe1998}
  Baraffe, I., Chabrier, G., Allard, F. \& Hauschildt, P. H. 1998,
  A\&A, 337, 403
\bibitem[\protect\citeauthoryear{Bedin et al.}{2003}]{bedin2003}
  Bedin, L. R., Piotto, G., King, I. R. \& Anderson, J. 2003, AJ, 126,
  247
\bibitem[\protect\citeauthoryear{Bedin et al.}{2009}]{bedin2009}
  Bedin, L. R., Salaris, M., Piotto, G. et al. 2009, ApJ, 697, 965
\bibitem[\protect\citeauthoryear{Bergeron et al.}{2011}]{bergeron}
  Bergeron et al. 2011, ApJ, 737, 28
\bibitem[\protect\citeauthoryear{Boudreault \&
    Lodieu}{2013}]{boudreault2013} Boudreault, S. \& Lodieu, N. 2013,
  MNRAS, 434, 142
\bibitem[\protect\citeauthoryear{Burgasser et al.}{2003}]{burgasser2003}
  Burgasser, A. J., Kirkpatrick, J. D., Burrows, A. et al. 2003, ApJ,
  592, 1186
\bibitem[\protect\citeauthoryear{Burgasser et al.}{2006}]{burgasser2006}
  Burgasser, A. J. \& Kirkpatrick, J. D. 2006, ApJ, 645, 1485
\bibitem[\protect\citeauthoryear{Burgasser et al.}{2009}]{burgasser2009}
  Burgasser, A. J., Witte, S., Helling, C. et al. 2009, ApJ, 697, 148
\bibitem[\protect\citeauthoryear{Burningham et
    al.}{2014}]{burningham2014} Burningham, B., Smith, L., Cardoso,
  C. V. et al. 2014, MNRAS, 440, 359
\bibitem[\protect\citeauthoryear{Burrows et al.}{1993}]{burrows1993}
  Burrows, A., Hubbard, W. B., Saumon, D. \& Lunine, J. I. 1993, ApJ,
  406, 158
\bibitem[\protect\citeauthoryear{Caiazzo et al.}{2017}]{caiazzo}
  Caiazzo, I., Heyl, J. S., Richer, H. \& Kalirai, J. 2017,
  arXiv:1702.0009
\bibitem[\protect\citeauthoryear{Casewell et al.}{2014}]{casewell2014}
  Casewell, S. L., Littlefair, S. P., Burleigh, M. R. \& Roy, M. 2014,
  MNRAS, 441, 2644
\bibitem[\protect\citeauthoryear{Cushing et al.}{2009}]{cushing2009}
  Cushing, M. C., Looper, D., Burgasser, A. J. et al. 2009, ApJ, 696, 986
\bibitem[\protect\citeauthoryear{Dieball et al.}{2016}]{dieball2016}
  Dieball, A., Bedin, L. R., Knigge, C. et al. 2016, ApJ, 817, 48
\bibitem[\protect\citeauthoryear{Dolphin}{2000}]{dolphin2000}
  Dolphin, A. E. 2000, PASP, 112, 1383
\bibitem[\protect\citeauthoryear{Girven et al.}{2011}]{girven2011}
  Girven, J., G\"ansicke, B. T., Steeghs, D. \& Koester, D. 2011,
  MNRAS, 417, 1210
\bibitem[\protect\citeauthoryear{Hansen et al.}{2004}]{hansen2004}
  Hansen, B. M. S., Richer, H. B., Fahlman, G. G. et al. 2004, ApJS, 155, 551
\bibitem[\protect\citeauthoryear{Holberg \& Bergeron}{2006}]{holberg}
  Holberg, J. B. \& Bergeron, P. 2006, AJ, 132, 1221
\bibitem[\protect\citeauthoryear{Hayashi \& Nakano}{1963}]{hayashi1963} Hayashi, C. \& Nakano, T. 1963, Progress of Theoretical Physics, 30, 460
\bibitem[\protect\citeauthoryear{Kaplan et al.}{2012}]{kaplan2012}
  Kaplan, M., Stramatellos, D. \& Whitworth, A. P. 2012, Ap\&SS, 341, 395
\bibitem[\protect\citeauthoryear{Kowalski \& Saumon}{2006}]{kowalski}
  Kowalski \& Saumon 2006, ApJ, 651, L137
\bibitem[\protect\citeauthoryear{Kulkarni}{1997}]{kulkarni1997}
  Kulkarni, S. R. 1997, Science, 276, 1350
\bibitem[\protect\citeauthoryear{Kumar}{1963}]{kumar1963} Kumar,
  S. S. 1963, ApJ 187, 1123
\bibitem[\protect\citeauthoryear{Lawrence et al.}{2007}]{lawrence2007}
  Lawrence, A., Warren, S. J., Almaini, O. et al. 2007, MNRAS, 379,
  1599
\bibitem[\protect\citeauthoryear{L\'epine et al.}{2004}]{lepine2004}
  L\'epine, S., Shara, M. M. \& Rich, R. M. 2004, ApJ, 602, 125
\bibitem[\protect\citeauthoryear{Nakajima et al.}{1995}]{nakajima1995}
  Nakajima, T., Oppenheimer, B. R., Kulkarni, S. R., Golimowski,
  D. A., Matthews, K. et al. 1995, Nature, 378, 463
\bibitem[\protect\citeauthoryear{Rebolo et al.}{1995}]{rebolo1995}
  Rebolo, R., Zapatero Osorio, M. R. \& Martín, E. L. 1995, Nature,
  377, 129
\bibitem[\protect\citeauthoryear{Richer et al.}{1997}]{richer1997}
  Richer, H. B., Fahlman, G. G., Ibata, R. A. et al. 1997, ApJ, 484, 741
\bibitem[\protect\citeauthoryear{Richer et al.}{2003}]{richer2003}
  Richer, H. B., Ibata, R., Fahlman, G. G. \& Huber, M. 2003, ApJL,
  597, 45
\bibitem[\protect\citeauthoryear{Richer et al.}{2004}]{richer2004}
  Richer, H. B., Fahlman, G. G., Brewer, J. et al. 2004, AJ, 127, 2771
\bibitem[\protect\citeauthoryear{Sanders}{1971}]{sanders} 
  Sanders, W. L. 1971, A\&A, 14, 226
\bibitem[\protect\citeauthoryear{Skrutskie et al.}{2006}]{skrutskie2006}
  Skrutskie, M., Cutri, R. M., Stiening, R. et al. 2006 AJ 131 1163
\bibitem[\protect\citeauthoryear{Stamatellos et al.}{2011}]{stamatellos2011}
  Stamatellos, D., Maury, A., Whitworth, A. \& Andr\'e, P. 2011,
  MNRAS, 413, 1787
\bibitem[\protect\citeauthoryear{Stetson}{1987}]{stetson} Stetson
  P. B. 1987, BAAS, 19, 745
\bibitem[\protect\citeauthoryear{Thies et al.}{2010}]{thies2010}
  Thies, I., Kroupa, P., Goodwin, S., Stamatellos, D. \& Whitworth,
  A. 2010, ApJ, 717, 577
\bibitem[\protect\citeauthoryear{Thies et al.}{2015}]{thies2015} Thies, I.,
  Pflamm-Altenburg, J., Kroupa, P. \& Marks, M. 2015, ApJ, 800, 72
\bibitem[\protect\citeauthoryear{Tremblay et al.}{2011}]{tremblay}
  Tremblay et al. 2011, ApJ, 730, 128
\bibitem[\protect\citeauthoryear{Troup et al.}{2016}]{troup2016}
  Troup, N. W., Nidever, D. L., De Lee, N. et al. 2016, AJ, 151, 85
\bibitem[\protect\citeauthoryear{York et al.}{2000}]{york2000} York, D. G.,
  Adelman, J., Anderson, J. E. et al. 2000, AJ, 120, 1579
\bibitem[\protect\citeauthoryear{Wright et al.}{2010}]{wright2010} Wright,
  E. L., Eisenhardt, P. R. M., Mainzer, A. K. et al. 2010, AJ, 140,
  1868
\end{thebibliography}
\end{document}